# Realizing the Data-Driven, Computational Discovery of Metal–Organic Framework Catalysts

Andrew S. Rosen, Justin M. Notestein,* Randall Q. Snurr*

Department of Chemical and Biological Engineering, Northwestern University, 2145 Sheridan Rd., Evanston, IL 60208 (USA)

*E-mail: j-notestein@northwestern.edu

*E-mail: snurr@northwestern.edu

**Highlights**:

- Big data approaches are needed to efficiently explore MOF catalyst space.
- High-throughput protocols have been developed to screen large MOF datasets.
- Machine learning is poised to accelerate MOF catalyst discovery.
- While challenges remain, there are many recent developments and new opportunities.

**Abstract**. Metal–organic frameworks (MOFs) have been widely investigated for challenging catalytic transformations due to their well-defined structures and high degree of synthetic tunability. These features, at least in principle, make MOFs ideally suited for a computational approach towards catalyst design and discovery. Nonetheless, the widespread use of data science and machine learning to accelerate the discovery of MOF catalysts has yet to be substantially realized. In this review, we provide an overview of recent work that sets the stage for future high-throughput computational screening and machine learning studies involving MOF catalysts. This is followed by a discussion of several challenges currently facing the broad adoption of data-centric approaches in MOF computational catalysis, and we share possible solutions that can help propel the field forward.

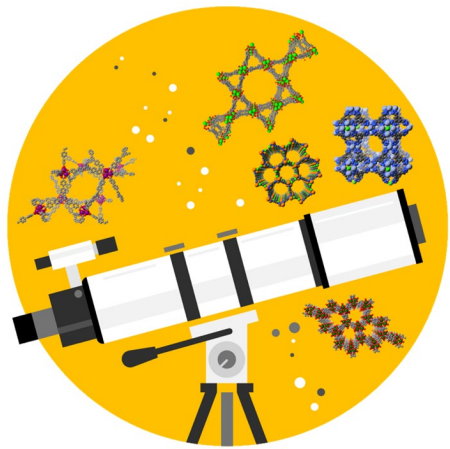

# Introduction

Metal–organic frameworks (MOFs) are a class of porous solids composed of metal ions or clusters connected by organic ligands. Due to their high degree of crystallinity and synthetic tunability, MOFs have been widely studied as novel catalytic materials for a variety of chemical transformations.[1] The atomically precise structures of most MOFs is especially advantageous for heterogeneous catalysis, as it makes it possible — at least in principle — to achieve fine-tuned control of the active site environment in ways that are typically reserved for molecular transition metal complexes.

Through different combinations of inorganic and organic building blocks, thousands of MOFs have been synthesized to date, and a virtually unlimited number can be proposed. With such a large number of possible MOFs that can be considered, it is rarely feasible to identify the optimal MOF for a given catalytic reaction based solely on intuition or trial-and-error experimental testing. Recent advances in data science, high-throughput computational screening, and machine learning (ML) represent a complementary route and can be used to rapidly identify promising MOFs from the vast combinatorial space of inorganic nodes, organic linkers, and topologies. Data-driven materials discovery, particularly based on high-throughput computational screening and ML, has been used to discover top-performing MOFs for numerous gas storage and separation processes.[2,3] Despite these successes — and the increasingly widespread use of data science techniques in the field of heterogeneous catalysis[4] — there remain extraordinarily few large-scale, data-driven studies of MOF-based catalysts.

In this review, we first provide a brief overview of recent studies that lay the groundwork for future big data and ML studies on MOF catalysts. We then discuss several challenges that must be overcome to fully enable a computational materials discovery pipeline for MOF catalysts. Drawing inspiration from successful uses of ML and materials informatics in adjacent application areas, we also highlight multiple opportunities to address the challenges currently facing data-driven MOF catalysis. Several of these themes are displayed in Figure 1.

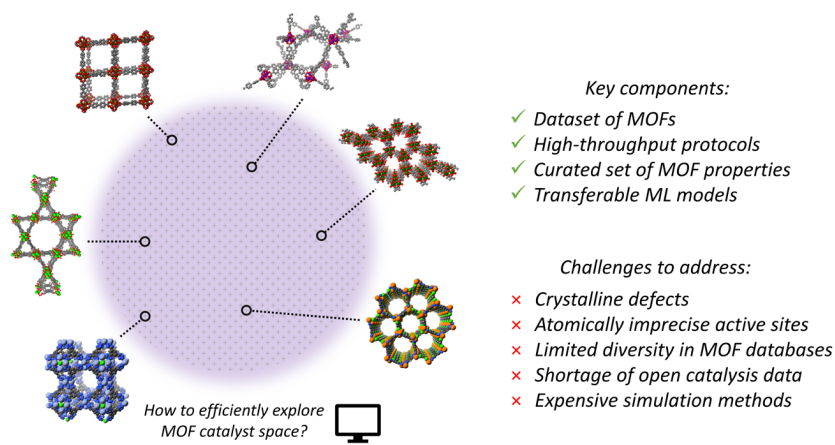

**Figure 1**. The enormous combinatorial space of inorganic nodes, organic linkers, and topologies results in a vast array of plausible MOF structures to consider for a given catalytic transformation. Several of the key components for a successful data-driven catalyst discovery approach are listed alongside some of the challenges currently facing the MOF computational catalysis community.

# Laying the Groundwork

In this section, we provide an overview of recent work involving high-throughput screening methods, MOF databases, and ML to predictively identify promising MOF catalysts and uncover important structure–property relationships. While high-throughput methods can involve experiments or theoretical calculations, we will focus predominantly on the latter given the state of the published catalysis literature. High-throughput catalytic experiments with MOFs are uncommon at the time of writing, with an important exception of several prior studies by Cohen and coworkers[5,6] who used high-throughput assays to identify promising MOF catalysts for nerve agent degradation. For a detailed review on high-throughput synthesis, characterization, and experimental testing of MOFs (and porous materials, such as zeolites), we refer the reader to an article by Clayson and coworkers.[7]

## *Small-to-Moderate-Scale Computational Screening*

Truly large-scale, high-throughput screening studies (e.g. examining several hundred or more materials) remain rare for MOF catalysis; nonetheless, several small-to-moderate-scale computational screening studies have been used to identify promising MOFs for ethylene dimerization,[8] light alkane oxidation,[9–11] nerve agent hydrolysis,[12] $CO_2$ fixation to epoxides,[13] electrocatalytic $CO_2$ reduction,[14] oxygen electrochemistry,[15] and alcohol dehydrogenation,[16] among other reactions.[17] A common theme amongst these studies is the use of density functional theory (DFT) calculations to predict catalytically relevant reaction energies and activation energies across a curated set of MOF families. Brønsted–Evans–Polanyi relationships, linear scaling relationships, and various structure–property relationships are then developed to gain chemical insight into the predicted reactivity trends.[17] As one example, Sours and Patel et al.[15] used DFT calculations to study ~30 porphyrin-based MOFs with different transition metal (TM) cations for the oxygen reduction reaction. The authors found that the presence of an oxophilic spectator (in this case, a nearby TM–OH group at an adjacent porphyrin linker) could stabilize adsorbed *OOH species via H-bonding interactions without substantially altering the corresponding *OH binding energy (Figure 2), potentially providing a way to improve the overall electrocatalytic performance.[17] Although the relatively small datasets in the aforementioned works (all less than 100 materials) preclude the use of modern data science techniques, they demonstrate the ability of computational methods to guide future experimental studies.

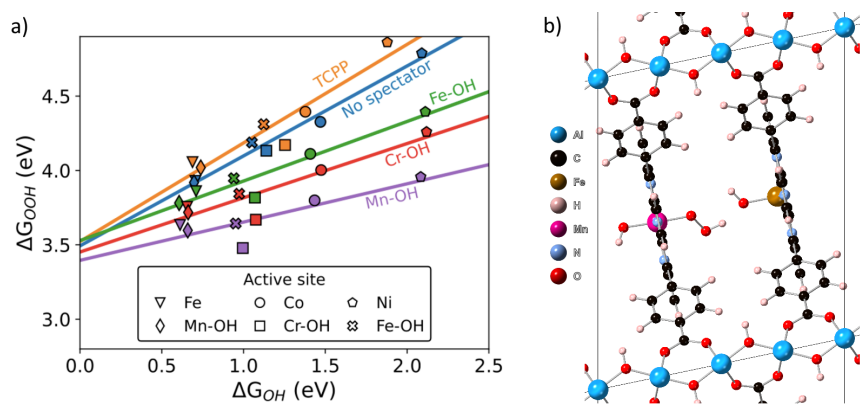

**Figure 2**. a) DFT-computed scaling relations for the *OOH and *OH binding free energies across several types of porphyrin-based catalysts. The orange data points labeled TCPP (TCPP = tetrakis(4-carboxyphenyl)-porphyrin)) refer to the isolated organic ligand with a single metal cation. The data points labeled "No spectator" refer to PMOF-Al (PMOF = porphyrin MOF) with only one type of transition metal cation in the porphyrin linkers (i.e. only the active site species). The remaining points refer to PMOF-Al with two types of transition metal environments — an active site species and oxophilic spectator — at adjacent porphyrin linkers. The active site species is indicated by the different

marker shapes. The green (Fe–OH), red (Cr–OH), and purple (Mn–OH) colors indicate the identity of the oxophilic spectator. Note that the Al species at the inorganic nodes are not directly involved in the proposed reactions. Figure reproduced with permission from Sours and Patel et al.[15] Copyright 2020 American Chemical Society. b) Example DFT-optimized structure for PMOF-Al with an Mn–OH active site (with bound *OOH species) and adjacent Fe–OH spectator (i.e. green diamond in Figure 2a). Atom color key: Al (teal), C (black), Fe (brown), H (cream), Mn (fuchsia), N (navy blue), O (red).

*Approaching High-Throughput Computational Screening*
The development of several curated databases of experimentally determined MOF crystal structures, such as the Computation-Ready, Experimental (CoRE) MOF Database[18] and Cambridge Structural Database (CSD) MOF Subset,[19] has made it possible to carry out large-scale computational screening studies. One of the first database-driven computational screening studies of MOF catalytic candidates was that of Vogiatzis et al.[20] who focused on the identification of MOFs that can activate $N_2O$ for the purpose of selectively oxidizing light alkanes. Starting from a dataset of ~5000 experimentally determined MOF crystal structures, the authors algorithmically identified a subset of MOFs with coordinatively unsaturated Fe sites, which are often studied for oxidation catalysis. Quantum-mechanical calculations based on DFT were then carried out in a conventional low-throughput mode to investigate the catalytic properties for seven MOFs of interest. Scaling up this approach further, Rosen et al.[21] developed a fully automated computational workflow that initializes the positions of small molecule adsorbates at metal binding sites and then carries out the necessary structure relaxations with periodic DFT to systematically identify promising MOF catalysts. In a separate study, the authors used this platform to screen a dataset of 60 MOF structures with a variety of inorganic nodes, organic linkers, and topologies to identify structure–reactivity relationships that can guide the discovery of MOFs for the catalytic activation of methane.[9]

Aside from thermal catalysis and electrocatalysis, MOFs have been widely investigated for their potential photocatalytic properties.[22] Arguably the most important properties of a MOF photocatalyst are its band gap and the energies of the corresponding band edges, which dictate the energy of light that can be absorbed and the redox properties, respectively. The energies of the valence band maximum and conduction band minimum must be aligned to a common reference, which can be done in an automated fashion by using the point in the MOF pore with the smallest variation in the electrostatic potential (i.e. the one closest to vacuum level).[23,24] With this protocol, Fumanal et al.[24] developed energy-based descriptors that allow for a high-throughput determination of the UV-vis light absorption capabilities and photo-redox properties of a dataset of MOFs. In a separate study by Fumanal et al.,[25] the authors developed a computational workflow to predict the charge separation and charge carrier mobility properties of MOFs in a high-throughput manner. These methodological advances pave the way for future computational screening studies of MOF photocatalysts.

*Machine Learning*
With regards to the use of ML for MOF catalyst discovery, there are currently few examples in the published literature. Arguably the closest study is that of Schweitzer and Archuleta et al.,[26] who constructed an ML model to determine the most important features dictating the DFT-computed binding energy of various atomic and small molecule adsorbates on simplified model systems of MOF-encapsulated nanoparticles. Li et al.[27] used experimentally reported turnover frequencies (TOFs) from 106 published experiments to train binary classification models that can predict whether a MOF might be active for carbon dioxide fixation; however, the low frequency of many of the one-hot encoded features (i.e. metals and linkers that appear only once in the dataset) and differences in how the TOFs were experimentally reported likely limit the predictive capabilities of the developed models. Although the number of ML studies specifically focused on MOF catalysis is limited, it should be noted that data-driven approaches in other application areas can be of significant value for catalyst discovery. For instance, ML models have been developed to predict the crystal morphology of MOFs from synthesis parameters[28] and the geometry of bound adsorbates directly from X-ray absorption near-edge structure spectra,[29] both of which can potentially aid the design of promising MOF catalysts.

## Challenges and Opportunities

Perhaps unsurprisingly, the primary factor currently holding back data-driven MOF catalysis is a lack of structured data from which to construct robust and generalizable models. While automated synthesis, characterization, and catalytic testing efforts have occasionally been undertaken,[7] the data (especially those involving "failed" experiments) are rarely made publicly available in a machine-readable format, and the total number of data points is often too small to build transferable ML models. From a computational perspective, high-throughput DFT calculations represent a natural way to generate large amounts of catalytically relevant data. In this section, we highlight some of the most important challenges facing data-driven computational catalysis with MOFs and suggest several ways that they can be addressed in future work.

### *High-Throughput DFT Property Databases for MOF Catalysis*

For catalytic reactions taking place on inorganic surfaces, there are multiple databases of DFT-computed adsorption energies (and occasionally, barrier heights) that can be used to develop data-driven models.[30,31] However, no such resource currently exists for MOFs. The closest analogue in the MOF field is the Quantum MOF (QMOF) Database,[32] which contains DFT-computed geometric and electronic structure properties for ~20,000 MOFs and coordination polymers. With algorithms to automatically position adsorbates at the metal sites of MOFs,[21] one can envision constructing a database of DFT-computed adsorption energies, particularly if the guest species are small and unlikely to induce complicated changes in spin state. Until such a resource becomes available, ML models based on transfer learning[33] may be able to take advantage of the large amounts of data available for other material classes (e.g. inorganic surfaces, transition metal complexes) to accurately predict MOF catalytic properties with a comparatively small amount of MOF-specific training data.

### *Computational Screening of Atomically Imprecise MOFs*

High-throughput computational screening of MOFs is also inherently tied to the quality of the underlying crystal structures being modeled. However, due to the presence of crystallographic disorder as well as the difficulty in resolving hydrogen atoms and charge-balancing ions, databases of experimentally synthesized MOF structures can rarely be screened without correcting or filtering out potentially problematic structures — a task that is far from trivial to automate. Fortunately, several approaches can be taken to increase the structural fidelity of a given MOF dataset. In addition to simple geometric heuristics that check for mis-bonded atoms and unlikely coordination environments,[34] it may be possible to use one of several ML models that can rapidly predict partial atomic charges in MOFs[35–37] to identify potential anomalies in a MOF crystal structure database. Jablonka et al.[38] demonstrated a somewhat similar concept with their ML-guided oxidation state prediction model, which was trained on user-reported data in the Cambridge Structural Database and was accurate enough to identify MOFs with erroneous oxidation state assignments. If developed, an ML model that could predict bond orders would also be able to efficiently flag structures with erroneously over- or under-coordinated framework atoms, generalizing what has been achieved with more empirical bond order heuristics in the literature.[34]

Similarly, while MOFs are often touted as having atomically precise structures, many of the most promising MOF catalysts in the published literature have undergone post-synthetic modifications (e.g. metal- or linker-exchange, linker functionalization, atomic layer deposition) that can introduce uncertainty in the structure of the active site environment.[39] Many MOFs are also known to exhibit defects in the crystal structure (e.g. missing nodes or linkers), which may greatly influence the catalytic properties of a given MOF but can be difficult to identify for the purposes of computational modeling. As one example, in the aforementioned screening study by Vogiatzis et al.,[20] Fe-BTT (BTT$^{3-}$ = 1,3,5-benzenetristetrazolate) was computationally identified as a promising MOF for oxidation catalysis when using $N_2O$ as an oxidant. Although the material was able to convert ethane to ethanol in experiments, spectroscopic measurements and catalyst cycling experiments led the authors to conclude that unidentified framework defects were likely responsible for the catalytic activity.[20] For large-scale computational screening studies that are likely to emerge in the future, it may prove beneficial to deliberately introduce carefully selected defects in databases of otherwise "pristine" MOF structures (e.g. using substructure find-and-replace methods[40]) to help ensure that the resulting data-driven models can accurately capture the complexities of experimentally synthesized MOF catalysts.

*Ensuring a Diverse MOF Dataset*

An alternate route to maximize the structural fidelity of a given MOF dataset is to use one of several hypothetical MOF construction tools[41–43] to computationally construct MOFs from a curated set of inorganic and organic building blocks. One cautionary aspect of this approach, however, is that databases of hypothetical MOFs can cover different regions of chemical space and/or lack the chemical diversity of experimentally synthesized MOF structures. Using an unsupervised dimensionality reduction technique, Moosavi et al.[44] showed that hypothetical MOF databases often lack significant diversity with regards to the inorganic nodes (Figure 3), which are of crucial importance for heterogeneous catalysis. Many of the hypothetical MOF databases that have been published to date are also tailored to gas storage and separation processes, where variations in pore geometry and linker length are likely to play a larger role than in many catalytic reactions, particularly those involving small molecules. To address potential issues related to structural diversity, a judicious selection of building blocks must be chosen, preferably before any computational catalyst screening studies are even carried out. This may be made easier with dimensionality reduction approaches like those shown in Figure 3 as well as larger databases of MOF building blocks, like that of Lee et al.[43] who amassed a dataset of over several hundred inorganic and organic building blocks, most of which are derived from experimentally reported MOF crystal structures.

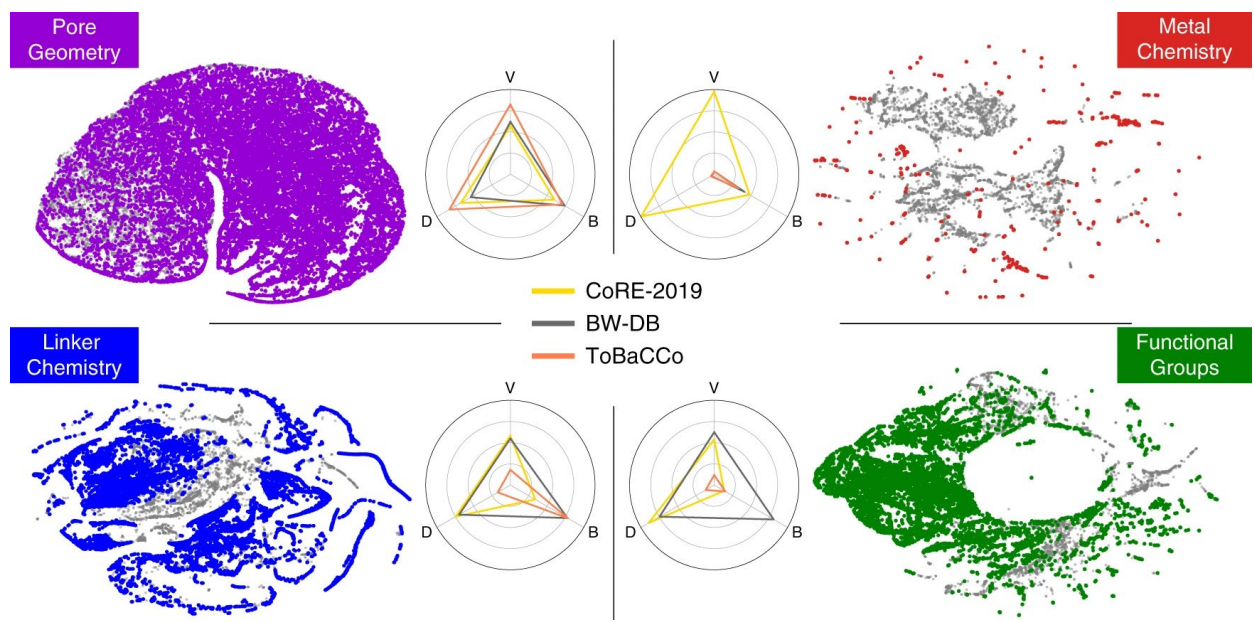

**Figure 3**. Dimensionality reduction based on the t-distributed stochastic neighbor method to highlight the pore geometry, metal chemistry, linker chemistry, and functional groups across three databases of MOF structures: the Computation-Ready, Experimental 2019 (CoRE-2019) Database, Boyd and Woo Database (BW-DB), and the Topologically Based Crystal Constructor (ToBaCCo) Database. The MOFs in the hypothetical databases (i.e. BW-DB, ToBaCCo) are shown as colored points overlaid on the entire MOF space from the three databases shown as gray points. The variety (V – how many bins across feature space are sampled), balance (B – how even the MOFs are distributed across feature space), and disparity (D – how spread out the MOFs are across feature space), which represent different diversity metrics, are shown for each database in the radar plots. The low degree of overlap with regards to metal chemistry between the hypothetical MOF databases and the database of experimentally synthesized MOF structures is likely to be especially relevant for screening MOF catalysts. Additionally, the two databases of hypothetical MOF structures exhibit varying degrees of overlap with the experimentally synthesized MOFs in the linker chemistry and functional group space, highlighting the need to carefully select the underlying inorganic and organic building blocks for a given application. The features used in the dimensionality reduction process are based on revised autocorrelation functions. Figure reproduced with permission from Moosavi et al.[44]

*Automated Construction of Representative Cluster Models*
When carrying out any theoretical study of MOF catalytic candidates, an important methodological decision is whether the system should be modeled as a crystalline lattice or as a truncated cluster model.[45] The former approach more closely resembles the true MOF structure, but this larger model system often limits the use of more accurate (and, therefore, more computationally expensive) theoretical methods. Carving a representative cluster model from the crystalline lattice takes advantage of the local nature of many catalytic reactions and results in a smaller model system that can often be studied with higher levels of theory. However, there are no automated approaches to determine the ideal size of the cluster model, where to terminate the cluster, how to cap the resulting dangling bonds, and which atoms should be kept rigid to appropriately mimic the bulk constraints of the crystalline framework. For high-throughput computational screening studies that involve a diverse range of MOF families, the fully periodic approach is often the only tractable option. Recently developed MOF deconstruction codes that can split up a MOF into its constituent building blocks[43,44,46] represent a first step towards automating the construction of MOF cluster models for high-throughput computational catalysis. Even if representative cluster models can be developed in a systematic manner, we note that such an approach will not be suitable for all catalytic reactions involving MOFs, such as reactions where confinement effects, long-range electrostatic interactions, or framework flexibility play an important role.

*Accelerated Identification of Ground State Magnetic Orderings*
The automated identification of ground state spin states for MOFs where open-shell character is possible, as is the case for most MOFs with 3$d$ transition metal cations,[45] represents yet another challenge for high-throughput computational MOF catalysis. The choice of spin state can drastically change the predicted reaction energies and barrier heights, and a separate quantum-mechanical calculation must be carried out for each plausible spin state to determine the lowest energy electron configuration. Even enumerating the different spin states can be a challenge in a high-throughput setting, particularly for MOFs that can exhibit internode and/or intranode magnetic exchange, redox reactions where the metal binding site and catalytic intermediates may exist in multiple possible redox states, and in cases where the metal oxidation state is not known *a priori*.

Reassuringly, ML models may be able to ease the computational burden of screening MOFs with magnetic character. As one example, the aforementioned ML model by Jablonka et al.[38] can be used to accurately assign oxidation states for each metal center in a MOF, which may help narrow down the list of possible spin multiplicities. There are also many recent studies to draw inspiration from in the homogeneous catalysis literature, such as the ML models developed by Taylor et al.[47] to predict the ground state spin states of transition metal complexes. High-throughput workflows have been developed to accurately and efficiently identify the ground state magnetic ordering in inorganic solids,[48] which may inspire the development of similar methods tailored to MOFs.

*Predicting MOF Stability and Synthesizability*

The previous sections have primarily focused on several of the hurdles facing both high-throughput computational screening and the development of data-driven models for MOF catalysis. However, it is also essential to know if a promising catalytic candidate will be able to be synthesized (if it is a newly proposed MOF) and, if it can be synthesized, whether it would be stable under the reaction conditions of interest. In a first major step towards determining whether a given MOF might be synthesizable, Anderson and Gómez-Gualdrón[49] carried out high-throughput molecular dynamics simulations to compute MOF free energies and were able to identify an energetic threshold, above which the likelihood of a MOF being readily synthesized can be considered too low for further consideration. Assuming a given MOF is synthetically plausible, Moosavi et al.[50] showed that the data from failed or inadequate synthesis attempts can be used to train machine learning models that guide the search for optimal synthesis conditions.

In terms of catalyst stability, there is a wide range of possible factors to consider, such as elevated temperatures, exposure to acidic or basic solutions, the presence of harsh oxidants, or a variety of other thermal and chemical stresses. Tackling one such stability criterion, Batra et al.[51] used a manually curated dataset of experimentally reported MOF–water interactions to train a machine learning-based classification model that can predict the hydrolytic stability of MOFs. To better understand mechanical stability, Moghadam et al.[52] used high-throughput molecular mechanics simulations to train an artificial neural network that can predict the bulk modulus of MOFs. Given the large number of published MOF studies that contain detailed characterization data, text mining and natural language processing techniques[53] have proven to be extremely valuable tools for the prediction of material properties, as has recently been demonstrated for the thermal and solvent-removal stability of MOFs.[54,55]

# Outlook

Looking into the future when automated theoretical calculations and experiments are likely to be more widespread in the field of MOF catalysis, an integrated materials discovery platform augmented by machine learning and data science can be envisioned (Figure 4). This concept has been referred to as "digital reticular chemistry"[56] and has the potential to revolutionize the MOF design and discovery process, including for the accelerated design of MOF catalysts tailored for challenging chemical transformations. In addition to the continued development of high-throughput methods,[7] we stress that it is essential for openly accessible data repositories to be developed in the area of MOF catalysis for this digital reticular chemistry future to be realized. Ideally, these MOF property databases should have standardized data formats and rich metadata,[57] in addition to abiding by the so-called FAIR guiding principles[58] (i.e. that the data is findable, accessible, interoperable, and reusable).

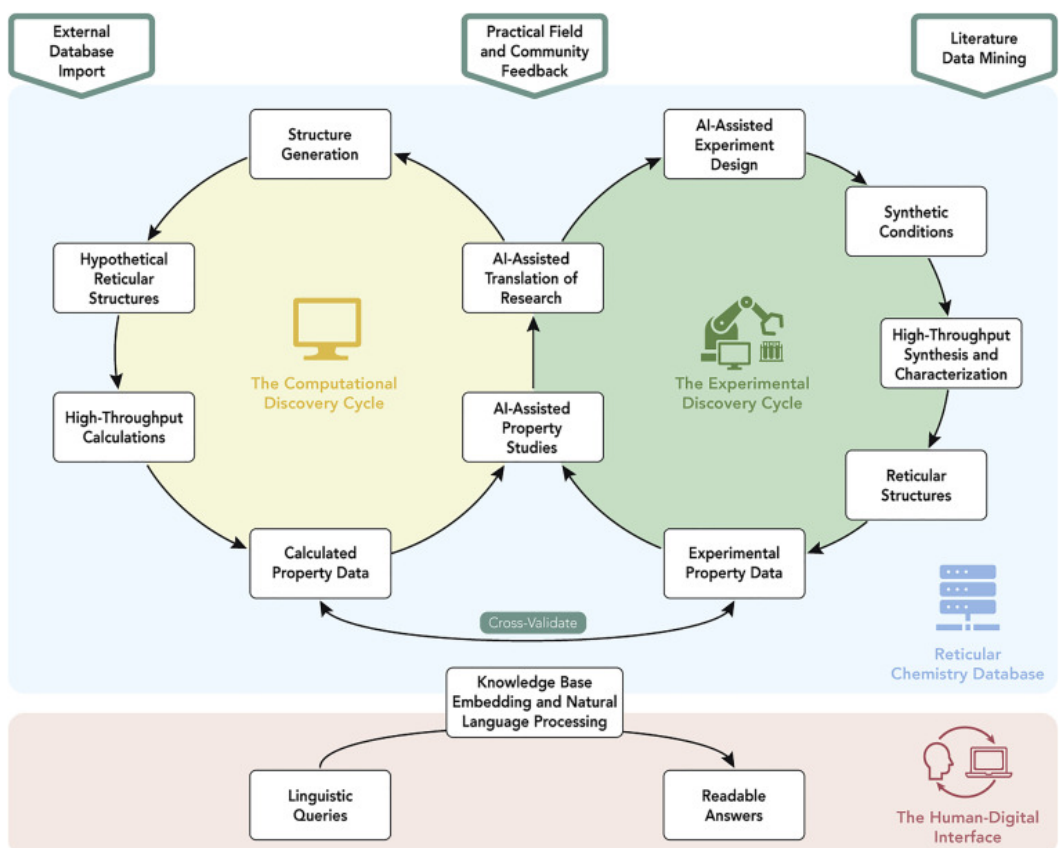

**Figure 4**. A schematic highlighting the proposed future of digital reticular chemistry. There are two main discovery cycles, one based on high-throughput theoretical calculations and the other based on high-throughput experiments. Both approaches are intrinsically connected, creating a closed feedback loop between the two. Artificial intelligence (AI) is used to help guide which theoretical calculations and experiments should be run. Data is stored in a reticular chemistry database for analysis and reuse. Domain knowledge, properties from external material databases, and data that can be text-mined from the published literature further guide the materials discovery pipeline. In addition to carrying out more detailed experiments on the most promising material candidates, the laboratory scientists and engineers help guide the workflow through queries enabled by natural language processing, in addition to more conventional means. Figure reproduced with permission from Lyu et al.[56] Copyright 2020 Elsevier Inc.

Collectively, the last several years have resulted in a gradual shift towards data-driven MOF catalysis, with several examples of small-to-moderate-scale screening studies published in the literature. Although a relative lack of carefully curated, catalytically relevant data currently limits the widespread adoption of ML in MOF catalysis research, we anticipate that this will rapidly change as recent advances in high-throughput screening approaches — both computational and experimental — become more widely adopted.

## Conflict of Interest Statement
The authors declare no conflict of interest.

## Acknowledgments

A.S.R. was supported by a fellowship award through the National Defense Science and Engineering Graduate (NDSEG) Fellowship Program, sponsored by the Air Force Research Laboratory (AFRL), the Office of Naval Research (ONR) and the Army Research Office (ARO). A.S.R. also acknowledges support from a Ryan Fellowship through the International Institute for Nanotechnology as well a Presidential Fellowship through The Graduate School at Northwestern University. The material in this work is supported by the Institute for Catalysis in Energy Processes (ICEP) via the U.S. Department of Energy, Office of Science, Office of Basic Energy Sciences (award number DE-FG02-03ER15457).